\documentstyle[amssymb,preprint,aps]{revtex}

\tightenlines

\begin{document}
\title{Spin-orbit scattering effect on critical current \\
in SFIFS tunnel structures }
\author{V.N.Krivoruchko, R.V.Petryuk}
\address{Donetsk Physics \& Technology Institute NAS of Ukraine, R.Luxemburg 72,\\
83114 Donetsk, Ukraine}
\date{\today}
\maketitle

\begin{abstract}
The spin-orbit scattering effect on critical current through
superconductor/ferromagnet ($SF$) bilayers separated by an insulator ($SFIFS$
tunnel junction) have been investigated for the case of absence of the
superconducting order parameter oscillations (thin $F$ layers). The analysis
is based on a microscopic theory for proximity coupled $\ SF$ bilayer for
different bilayer parameters (boundary transparency, proximity effect
strength, relative orientation and value of the $F$ layers exchange field) .
We find that the spin-orbit scattering considerably modifies dc Josephson
current in $SFIFS$ tunnel junction. In contrast to a simple physical
picture, the reduction of the exchange field effects is nonlinear in
character getting its maximum in the field's region where the critical
current enhancement or the transition to the $\pi $-state take place. Hence,
for understanding the various experimental results on tunnel structures with
thin $F$ layers, the coupled effects of the exchange interaction and
spin-orbit scattering must be considered.
\end{abstract}

\section{INTRODUCTION}

Nowadays progress in nanotechnology made it possible to produce
nanostructures with new physical phenomena. This has led to renewed
attention to hybrid systems consisting of superconductor ($S$) and
ferromagnetic ($F$) metals, and displaying rich and elegant physics and
having potential applications. Transport properties of $SF$ structures with
artificial geometry have turned out to be quite unusual. These have been
treated by several authors [1-9], and the obtained results show that in
ferromagnet the Cooper pair potential not only exponentially decays but also
has an oscillatory character; i.e., in a ferromagnet the density of Cooper
pairs is spatially inhomogeneous and the superconducting\ order parameter
contains nodes where the phase changes by $\pi $. This causes the exchange
field dependence of the Josephson coupling energy, and if the exchange
energy $H_{exc}$ in magnetic layer exceeds a certain value,\ a crossover
from $0$-phase to $\pi $-phase superconductivity takes place. The phenomenon
has been theoretically described for $SFS$ weak links with thick $F$ layer
[1-5]. The crossover to the $\pi $-state even in the absence of the order
parameter oscillations in thin $F$ layer was also predicted for $SFIFS$
tunnel junctions (where $I$ is an insulator) with parallel alignment of
layer's magnetization [6]. For an antiparallel orientation, the possibility
of the critical current enhancement by the exchange interaction in $SFIFS$
junctions with thin $F$ layers was discussed both for small [7] and bulk
[8]\ $S$\ layer thicknesses. Experimentally, the $\pi $-phase
superconductivity has been observed for $SFS$\ weak links [9] and $SIFS$
[10] and $SNFNS$\ [11] tunnel junctions with thick $F$ layers, while the
enhancement of the dc Josephson current has not been detected until\ now.

Some features should be taken into account for theory to be adjustable for
experimental results on $SF$ structures, and the spin-flip processes are
among the important ones. These processes can be induced, e.g., by the
spin-orbit scattering centers presented in the film; other important source
of the spin-flip processes for nanoscale hybrid structures is a\ strong
electric field arising near metal-metal boundaries [12]. The basic physics
behind spatial\ oscillations of induced superconductivity in $SF$ sandwiches
with spin-orbit scattering has been recently discussed in [3,13].\ As is
known, in the presence of spin-orbit scattering the electron spin is no
longer a good quantum number and the electron will change its spin state
during a characteristic time $\tau _{SO}$ , while the Cooper pair will mix
with its spin-exchanged counterpartner. This causes a pair to ''feel'' an
exchange field which changes sign at a rate proportional to $1/\tau _{SO}$ ,
decreasing the average ferromagnetic field experienced by the pair. That
means, that the spin-flip processes not only modify the oscillation length
but also lead to an extra decay of the Cooper pair potential and, at a
critical strength, these scattering processes can completely suppress the $%
\pi $-phase superconductivity.

The scenario of the $0-\pi $-phase transition, as well as the mechanisms of
critical current enhancement, in the limit of thin $F$ layers differ from
those for thick $F$ layers. However, the basic physics behind spin-orbit
scattering effect for the $SF$ sandwiches where there is no induced order
parameter oscillations, has not been discussed till now. The purpose of this
paper is to clarify the mechanisms of influence of spin-orbit scattering on
critical current in $SFIFS$ tunnel structures with thin $F$ layers. Our
analysis is based on the extension of calculations [6,8,14], so as to
include the effects of the spin-orbit scattering. Namely, we study the
tunnel junction with leads formed by the proximity coupled $SF$ bilayers of
a massive $s$-wave superconductor and a thin $F$ metals, when the spin-orbit
scattering processes take place in ferromagnetic layers. The microscopic
model of the proximity effect for $SF$ bilayer is employed to discuss the
case of arbitrary value of $SF$ boundary transparency, ferromagnetic
exchange field, and proximity effect strength (Sec.II). Critical current of
symmetrical $SFIFS$ junction is discussed in Sec.III where analytical
solutions have been obtained for some particular cases. As we shall see,
spin-flip scattering plays a major role in transport properties of
superconductor-ferromagnet structures with thin $F$ layers and should be
considered for understanding the various experimental results. We close with
a Conclusion.

\section{ SUPERCONDUCTIVITY OF SF BILAYER\ WITH SPIN-ORBIT SCATTERING}

We consider the case when\ both metals are in the dirty limit condition $\xi
_{S,F}\geq l_{S,F}$ , and\ the $S$ layer is thick, $d_{S}\gg \xi _{S}$,
while the $F$ layer is thin, $d_{F}\ll \min (\xi _{F},\sqrt{D_{F}/2\pi T_{C}}%
)$. Here $\xi _{S}=(D_{S}/2\pi T_{C})^{1/2}$ and $\xi
_{F}=(D_{F}/2H_{exc})^{1/2}$ are the coherence lengths of the $S$ and $F$
metals; $l_{S}$ and $l_{F}$ are the electron mean free paths, $d_{S,F}$ are
the thicknesses, and $D_{S,F}$ are the diffusion coefficients of the $S$ and 
$F$ metals, respectively. These conditions make it possible to neglect the
reduction of the critical temperature of the $SF$ bilayer compared to that
of the bulk $S$ metal, and to imply constant superconducting properties
through $F$ layer thickness. Throughout this work the $F$ layer will be
treated as a single domain film with the spin-orbit scattering centers,
while there are no such centers in the $S$ layer. Then, the superconducting
properties of the $SF$ bilayer are described by the Usadel equations [15].
The latter can be written as (the domain $x\geq 0$ is occupied by the $S$
metal and ${\it x}<0$ by the $F$ metal):

\begin{equation}
\Phi _{S\sigma }=\Delta _{S}+\xi _{S}{}^{2}\frac{\pi T_{C}}{\omega
G_{S\sigma }}[G_{S\sigma }{}^{2}\Phi _{S\sigma }^{\prime }]^{\prime }\text{ }%
,  \label{ref1}
\end{equation}
\begin{equation}
\Phi _{F\pm }=\xi ^{2}\frac{\pi T_{C}}{\tilde{\omega}_{\pm }G_{F\pm }}%
[G_{F\pm }^{2}\Phi _{F\pm }^{\prime }]^{\prime }+\alpha _{SO}G_{F\mp }(\frac{%
\Phi _{F\mp }}{\tilde{\omega}_{\mp }}-\frac{\Phi _{F\pm }}{\tilde{\omega}%
_{\pm }})
\end{equation}
Here $\tilde{\omega}_{\sigma }=\omega +i\sigma H_{exc}$, \ $\sigma =\pm $ ,
and $\omega \equiv \omega _{n}=\pi T(2n+1)$, $\ n=\pm 1,\pm 2,...$ is
Matsubara frequency; $\alpha _{SO}=2/3\tau _{SO}$\ and $\tau _{SO}$ is the
spin-orbit scattering time; the pair potential $\Delta _{S}$ is determined
by the usual self-consistency relations (see, e.g. [16]). We define the $x$
axis as perpendicular to the film surfaces and the prime denotes
differentiation with respect to a coordinate ${\it x}$. The modified Usadel
functions [17] $\Phi _{S\sigma }=\omega F_{S\sigma }/G_{S\sigma }$, $\Phi
_{F\sigma }=\tilde{\omega}_{\sigma }F_{F\sigma }/G_{F\sigma }$ , where $%
G_{F,S}$ and $F_{F,S}$ are normal and anomalous Green's functions for the $F$
and $S$ material, respectively, are introduced\ to take into account the
normalized confinement of the Green's functions $G^{2}+F\tilde{F}=1$\ ;
here\ $\tilde{F}(\omega ,H_{exc})=F^{\ast }(\omega ,-H_{exc})$ , and $\tilde{%
\Phi}_{S\sigma }=\omega \tilde{F}_{S\sigma }/G_{S\sigma }$ (see below).\ We
also assume that for a nonsuperconducting $F$ metal the bare value of the
order parameter $\Delta _{F}=0$\ , however, the Cooper pair correlation
function $F_{F}\neq 0$ due to the proximity effect. Eqs. (2) for the $F$
metal are the generalization of the Usadel equations for the\ case when the
spin-orbit scattering is present (for more details see Ref. [18]). At
temperatures close to $T_{C}$ , when $G_{S(F)\sigma }\approx 1$ and $\Phi
_{S(F)\sigma }/\tilde{\omega}_{\sigma }\approx F_{S(F)\sigma }$ , Eqs. (2)
have been simplified getting the form of Eqs. (26) of Ref. [3]. If the
spin-orbit processes are absent$\ (\tau _{SO}\rightarrow \infty )$\ , the
spin ''up'' and ''down'' subbands do not mix with each other and the
equation (2) obtains the usual form (see, e.g., [6,14]).

Eqs. (1), (2) should be supplemented with the boundary conditions. In the
bulk of the $S$ metal the pair potential is equal to the $BCS$ order
parameter $\Delta _{0}(T)$\ at the temperature $T$ : $\Phi _{S}(\infty
)=\Delta _{S}(\infty )=\Delta _{0}(T)$, while at the free (dielectric)
interface of the $F$ metal:\ $\Phi _{F\pm }^{/}(-d_{F})=0$. Assuming that
there are no spin-flip processes at the $SF$ interface, the boundary
conditions at this interface (see Ref. [14] for details) can easily be
generalized for the case of two fermionic subband in the form :

\begin{equation}
\gamma \xi G_{F\pm }^{2}\Phi _{F\pm }^{/}/\tilde{\omega}_{\pm }=\xi
_{S}G_{S\pm }^{2}\Phi _{S\pm }^{/}/\omega ,
\end{equation}
\begin{equation}
\gamma _{BF}\xi G_{F\pm }\Phi _{F\pm }^{/}=\tilde{\omega}_{\pm }G_{S\pm
}(\Phi _{S\pm }/\omega -\Phi _{F\pm }/\tilde{\omega}_{\pm })
\end{equation}
The parameters $\gamma $ and $\gamma _{BF}$ involved in these relations are
given by $\gamma =\rho _{S}\xi _{S}/\rho _{F}\xi $ , $\gamma
_{BF}=R_{B}/\rho _{F}\xi $ , where $\rho _{S},_{F}$ are the normal state
resistances of the $S$ and $F$ metals, $R_{B}$ is the product of $SF$
boundary resistance and its area. In Eqs. (1) and (3), and below we have
used the effective coherence length $\xi =(D_{F}/2\pi T_{C})^{1/2}$ thus
providing the regular crossover to both limits $T_{C}\gg H_{exc}\rightarrow
0 $ and $H_{exc}\gg T_{C}$. The relation (3) provides the continuity of the
supercurrent flowing through the $SF$ interface at any value of the
interface transparency for spin ''up'' and ''down'' fermionic subband
separately, while the condition (4) accounts for the quality of the electric
contact. An additional physical approximation for the relations (3) and (4)
to be valid is the assumption that spin discrimination by the interface is
unimportant, i.e., the interface parameters involved are the same for both
spin subband. Generalization of Eqs. (3) and (4) to the $SF$ interface with
different transmission probabilities for ''up'' and ''down'' spin
quasiparticles is straightforward, however the case we consider contains all
the new physics we are interested in and is simpler. In such ferromagnetic
metals as $Ni$, $Gd$, etc. the polarization of the electrons at low
temperatures is not more than 10\%, and one can expect the model under
consideration reflects the transport properties of these ferromagnets hybrid
structures.

Due to a small thickness of the $F$ metal the proximity effect problem can
be reduced to the boundary problem for the $S$ layer and a relation for
determining $\Phi _{F\pm }$ at $x=0$ . There are three parameters which
enter the model: $\gamma _{M}=\gamma d_{F}/\xi $ ,\ $\gamma _{B}=\gamma
_{BF}d_{F}/\xi $ and the energy of the exchange field $H_{exc}$\ . Using the
system of equations (3) and (4), one can obtain the equations determining
the unknown value of the functions $\Phi _{F\pm }(0)$ and boundary
conditions\ for $\Phi _{S\pm }$ . Due to these boundary conditions all the
equations for the functions $\Phi _{S\pm }$ and $\Phi _{F\pm }$ are coupled.
In the general case, the problem needs self-consistent numerical
calculations (similar to those as, e.g. , in Ref. [19]). Here, however, we
will not discuss the quantitative solution, but pay attention to the
qualitative one to consider new physics we are interested in.

We will hereinafter assume that $\alpha _{SO}\ll 1$ and$\ d_{F}/\xi \ll 1$
and solve differential equations by an iteration procedure finding the
corrections to $\Phi _{F\pm }(x)$ and$\ \Phi _{S\pm }(x)$ in small
parameters $d_{F}/\xi $ and $\alpha _{SO}$. We also restrict ourselves to
the quite realistic experimental case of a bilayer with a weak proximity
effect and low transparency of the $SF$ boundary, i.e., $\gamma _{M}\ll 1$
and $\gamma _{B}\gtrsim 1$ . Then the problem is simplified (see Ref. [18]
for details) and reduced to the Usadel Eqs. (1) for the $S$ layer with
boundary conditions (3), while Eq. (4) reduces to

\begin{equation}
\xi _{S}G_{S\pm }\Phi _{S\pm }^{/}|_{x=0}\approx \bar{\gamma}_{M}\tilde{%
\omega}_{\pm }\frac{\Phi _{S\pm }}{\pi T_{C}A_{\pm }}\{1\mp 2\alpha _{SO}%
\frac{iH_{exc}}{\tilde{\omega}_{+}\tilde{\omega}_{-}}\frac{G_{S}}{\omega }(%
\frac{\omega }{\gamma _{B}\tilde{\omega}_{\pm }}\pi T_{C}+\frac{G_{S}}{%
\omega }\frac{\Delta _{0}^{2}}{A_{\pm }^{2}})\}|_{x=0},
\end{equation}
where we abbreviated $A_{\pm }=[1+2\bar{\gamma}_{B}G_{S}\tilde{\omega}_{\pm
}+\bar{\gamma}_{B}^{2}\tilde{\omega}_{\pm }^{2}]^{1/2}$ , $G_{S}=\omega
/(\omega ^{2}+\Delta _{0}^{2})^{1/2}$ and $\bar{\gamma}_{B(M)}\equiv \gamma
_{B(M)}/\pi T_{C}$ (functions that are multiplied by $\alpha _{SO}$ have
been used in the limit $\alpha _{SO}\rightarrow 0$). Now, the equation for $%
F $ layer function $\Phi _{F\pm }(0)$ has the form:

\begin{equation}
\Phi _{F\pm }(0)\approx G_{S\pm }\Phi _{S\pm }\{1\pm 2\alpha _{SO}\frac{%
iH_{exc}}{\tilde{\omega}_{+}\tilde{\omega}_{-}}\}/[\omega (\bar{\gamma}%
_{B}+G_{S\pm }/\tilde{\omega}_{\pm }]|_{x=0}\ 
\end{equation}
In zeroth approximation in $\gamma _{M}$ the $\Phi _{S\pm }^{/}(0)=0$. So,
we can neglect the suppression of superconductivity in the $S$ layer
assuming that $\Phi _{S\pm }(x)$ is spatially homogeneous: $\Phi _{S\pm
}(x)=\Delta _{S}(x)=\Delta _{0}(T)$ . In the next order in $\gamma _{M}$ ,
by linearizing the Usadel equations for the $\Phi _{S\pm }(x)$ and making
use of the relation (3), the general solution of the linearized equation (1)
is given by

\begin{equation}
\Phi _{S\pm }(x)=\Delta _{0}\{1-C_{\pm }\exp (-\beta x/\xi _{S})\}
\end{equation}
where $\beta ^{2}=(\omega ^{2}+\Delta _{0}^{2})^{1/2}/\pi T_{C}$.
Substituting this solutions into the boundary relations (5), we get for $%
C_{\pm }$\ :

\begin{equation}
C_{\pm }\approx \frac{\bar{\gamma}_{M}\beta \tilde{\omega}_{\pm }}{\bar{%
\gamma}_{M}\beta \tilde{\omega}_{\pm }+\omega A_{\pm }}\{1\mp 2\alpha _{SO}%
\frac{iH_{exc}}{\tilde{\omega}_{+}\tilde{\omega}_{-}}(\frac{G_{S}}{\bar{%
\gamma}_{B}\tilde{\omega}_{\pm }}+\frac{G_{S}^{2}}{\omega ^{2}}\frac{\Delta
_{0}^{2}}{A_{\pm }^{2}})\}
\end{equation}
Using Eqs. (7) for $x=0$ and relations (6), one can find expressions for $%
\Phi _{F\pm }(\omega ,0)$. However we will not show here these expressions
because of their cumbersome structure. For $H_{exc}\rightarrow 0$\ the
quantity $\tilde{\omega}_{\pm }\rightarrow \omega $\ ,\ and solution (7),
(8) reproduces the results obtained in Ref. [16]\ for an $SN$ bilayer. If $%
\tau _{SO}\rightarrow \infty $, the expressions restore earlier results for $%
SF$ bilayer (see Refs.\ [6,14]).

Let us give some comments on the results obtained. For the structure under
consideration, one can expect a kind of induced magnetic properties for $S$
layer, that are the result of the effect similar to the superconducting
proximity effect [20]. In accordance with Eqs. (7), (8), we see that due to
proximity induced magnetic correlations even in the $S$ layer fermionic
symmetry of subbands has been lost and Cooper pair mixes with its
spin-exchanged counterpartner.

\section{CRITICAL CURRENT OF $SFIFS$ TUNNEL JUNCTION}

We assume that both banks of the Josephson $SFIFS$ tunnel junction are
formed by equivalent $SF$ bilayers, and the transparency of the insulating
layer is small enough to neglect the effect of the tunnel current on the
superconducting state of the electrons. The plane $SF$ boundary can have
arbitrary finite transparency, but it is large compared to the transparency
of the junction barrier. The transverse dimensions of the junction are
supposed to be much less then the Josephson penetration depth, so all
quantities depend only on a single co-ordinate $x$ normal to the interface
surface of the materials. Using the above obtained results we investigate
the influence of the spin-flip scattering on critical Josephson current in $%
SFIFS$ tunnel junction. The critical current of the ($SF)_{L}I(FS)_{R}$
tunnel contact can be written in the form (see, e.g. Ref. [6]):

\begin{equation}
j_{C}=(eR_{N}/2\pi T_{C})I_{C}=\frac{T}{T_{C}}%
\mathop{\rm Re}%
\sum_{\omega >0,\sigma =\pm }\frac{G_{F\sigma }\Phi _{F\sigma }}{\tilde{%
\omega}_{\sigma }}|_{L}\frac{G_{F\sigma }\Phi _{F\sigma }}{\tilde{\omega}%
_{\sigma }}|_{R}
\end{equation}
where $R_{N}$ is the resistance of the contact in the normal state; the
subscript $L$ ($R$) labels quantities referring to the left (right) bank and
the sign of the exchange field depends on mutual orientation of the bank
magnetizations.

\subsection{Parallel orientation of the layer's magnetizations{\it \ }}

For parallel alignment of the layer's magnetizations, i.e., with $\tilde{%
\omega}_{L}=\tilde{\omega}_{R}$ , the expression for critical current reads:

\begin{eqnarray}
j_{C}^{FM} &=&\frac{T}{T_{C}}%
\mathop{\rm Re}%
\sum_{\omega >0,\sigma =\pm }\frac{G_{S\sigma }^{2}\Phi _{S\sigma }^{2}}{%
\omega ^{2}}\{1+4\sigma \alpha _{SO}\gamma _{B}\frac{iH_{exc}}{\tilde{\omega}%
_{+}\tilde{\omega}_{-}}\}\times  \nonumber \\
&&\{1+2\bar{\gamma}_{B}G_{S\sigma }\tilde{\omega}_{\sigma }+\bar{\gamma}%
_{B}^{2}\tilde{\omega}_{\sigma }^{2}+\sigma 4\alpha _{SO}\bar{\gamma}_{B}%
\frac{iH_{exc}}{\tilde{\omega}_{+}\tilde{\omega}_{-}}\frac{G_{S\sigma
}^{2}\Phi _{S\sigma }\tilde{\Phi}_{S\sigma }}{\omega ^{2}}\}^{-1}
\end{eqnarray}
We begin with an analytical consideration for the case of a vanishing
effective pair breaking parameter near the $SF$ boundary, $\gamma _{M}=0$;
i.e., the influence of the $F$ layer on the superconducting properties of
the $S$ metal can be neglected and the order parameter in the $S$ region is
spatially homogeneous: $\Phi _{S\pm }(x)=\Delta _{0}(T)$\ (see, Eqs.\ (7)
and (8)).\ For the amplitude of the Josephson current we than obtained

\begin{eqnarray}
j_{C}^{FM} &\approx &2\frac{T}{T_{C}}\sum_{\omega >0}\frac{\Delta _{0}^{2}}{%
\Delta _{0}^{2}+\omega ^{2}}\{1+2\bar{\gamma}_{B}\omega G_{S}+\bar{\gamma}%
_{B}^{2}(\omega ^{2}-H_{exc}^{2})+8\alpha _{SO}\bar{\gamma}_{B}^{2}\frac{%
\omega H_{exc}^{2}}{\omega ^{2}+H_{exc}^{2}}\}\times  \nonumber \\
&&\{[1+2\bar{\gamma}_{B}\omega G_{S}+\bar{\gamma}_{B}^{2}(\omega
^{2}-H_{exc}^{2})]^{2}+4H_{exc}^{2}\bar{\gamma}_{B}^{2}(G_{S}+\bar{\gamma}%
_{B}\omega )^{2}\}^{-1}
\end{eqnarray}
If $H_{exc}\rightarrow 0$ the expression (11) restores the result for $SNINS$
junction (see, e.g., Eq.(28a) of Ref.[17] for parameters value under
consideration). As is seen from this expression, for large enough $H_{exc}$
, the supercurrent changes its sign, i.e., with increasing magnetic energy
the junction crosses over from $0$-phase type to $\pi $-phase type
superconductivity. However, the spin-flip processes exert influence upon
junction's tendency to set in the $\pi $-phase state.

On Fig. 1 we plot a\ family of the Josephson current amplitude (11) as
function of $H_{exc}$ when $\gamma _{B}=2$ , $\gamma _{M}=0$ for various
values of the spin-orbit scattering intensity $\alpha _{SO}$ . Here we also
show the function $\delta j_{C}^{FM}=j_{C}^{FM}(\alpha
_{SO})-j_{C}^{FM}(\alpha _{SO}=0)$ .The main feature here is that the
increase of the exchange energy pulls the $SFIFS$ junction to $\pi $-state.
The new result of this figure is that, as intensity of spin-orbit scattering
processes increases, the critical current amplitude, $j_{C}^{FM}(\alpha
_{SO}),$ for the $\pi $-state decreases. The spin-orbit scattering reduces
the effect of exchange field, and this reduction has nonlinear character
with its maximum in the region where transition to the $\pi $-state take
place.

A more realistic case is shown on Figs. 2, where we plot the results of
numerical calculations of the critical current, Eq. (10), for the tunnel
junction when suppression of the $S$\ layer superconductivity occurs due to
proximity effect $(\gamma _{M}\neq 0)$. The function $\delta j_{C}^{FM}$ has
also been shown. Again, the main result of these calculations is that the
spin-flip processes can sizably reduce the $SFIFS$ junction tendency to $\pi 
$-phase state. In contrast to a simple physical picture, the suppression has
nonlinear behavior getting its maximum near $H_{exc}$ region of cross over
from $0$-type to $\pi $-type superconductivity. While our results have been
obtained for the $SF$ bilayers with\ a weak proximity effect and low
boundary transparency, qualitatively the conclusions are valid for arbitrary 
$\gamma _{B}$ and $\gamma _{M}$ values.

\subsection{Antiparallel orientation of the layer's magnetizations}

To be definite, we took $\tilde{\omega}_{L}=\omega +iH_{exc}$, $\tilde{\omega%
}_{R}=\omega -iH_{exc}$. Now the critical current of the $(SF)_{L}I(FS)_{R}$
tunnel contact can be given in the form

\begin{eqnarray}
j_{C}^{AF} &=&2\frac{T}{T_{C}}%
\mathop{\rm Re}%
\sum_{\omega >0}\frac{G_{S+}\Phi _{S+}G_{S-}\Phi _{S-}}{\omega ^{2}}%
\{(G_{S+}+\bar{\gamma}_{B}\tilde{\omega}_{+})^{2}+\frac{G_{S+}^{2}\Phi _{S+}%
\tilde{\Phi}_{S+}}{\omega ^{2}}(1+4\alpha _{SO}\frac{iH_{exc}}{\tilde{\omega}%
_{+}\tilde{\omega}_{-}})]\}^{-1}  \nonumber \\
&&\times \{(G_{S-}+\bar{\gamma}_{B}\tilde{\omega}_{-})^{2}+\frac{%
G_{S-}^{2}\Phi _{S-}\tilde{\Phi}_{S-}}{\omega ^{2}}(1-4\alpha _{SO}\frac{%
iH_{exc}}{\tilde{\omega}_{+}\tilde{\omega}_{-}})\}^{-1}
\end{eqnarray}
Fist we assume that one can neglect the suppression of the superconductivity
in the $S$ layer ($\gamma _{M}=0$). Then, after simple but cumbersome
algebra we have for the critical current 
\begin{eqnarray}
j_{C}^{AF} &\approx &2\frac{T}{T_{C}}\sum_{\omega >0}%
{\displaystyle{\Delta _{0}^{2} \over \Delta ^{2}+\omega _{0}^{2}}}%
\{1-8\alpha _{SO}%
{\displaystyle{\Delta _{0}^{2} \over \Delta _{0}^{2}+\omega ^{2}}}%
\frac{\omega H_{exc}^{2}}{\bar{\gamma}_{B}^{2}(\omega ^{2}+H_{exc}^{2})}%
\}\{[1+2\bar{\gamma}_{B}\omega G_{S}+\bar{\gamma}_{B}^{2}(\omega
^{2}-H_{exc}^{2})]^{2}+  \nonumber \\
&&4H_{exc}^{2}\bar{\gamma}_{B}^{2}(G_{S}+\bar{\gamma}_{B}\omega
)^{2}\}^{-1/2}
\end{eqnarray}
For $H_{exc}\rightarrow 0$ expression (13) restores the result for $SNINS$
junction (Eq. (28a) [17]). One can see, that for $\omega <H_{exc}$\ the
expression in figured brackets is low than one. As a result, the $0$-phase
state across the contact has been preserved, and in some interval of the
exchange field energy the enhancement of dc Josephson current occurs. The
spin-orbit processes exert upon this behavior, reducing the current
enhancement.

On Fig. 3 we show a family of the Josephson current amplitude (13) as
function of $H_{exc}$ when $\gamma _{B}=2$ , $\gamma _{M}=0$ for various
values of the spin-orbit scattering intensity $\alpha _{SO}$ . Here we also
plot the function $\delta j_{C}^{AF}=j_{C}^{AF}(\alpha
_{SO})-j_{C}^{AF}(\alpha _{SO}=0)$ .The main and unusual feature here is
that the increase of the exchange energy enhances the $SFIFS$ junction
critical current. The new result of this figure is that, as the intensity of
spin-orbit scattering processes increases, the tendency of the $%
j_{C}^{AF}(\alpha _{SO})$ to enhance decreases. The spin-orbit scattering
reduces the effect of exchange field, and this reduction is nonlinear in
character with its maximum in the most interesting for experimentalists
region. More general cases are shown on Fig. 4, where we illustrate the
results of numerical calculations of the critical current, Eq. (12), taking
into account a suppression of the $S$ layer superconductivity by the
proximity effect ($\gamma _{M}\neq 0)$. The function $\delta j_{C}^{AF}$ has
also been shown. One can see again, that the spin-orbit suppression has
nonlinear behavior getting its maximum in the most interesting for
experimental investigation region of $H_{exc}$ . Qualitatively, these
results are valid for arbitrary values of $\gamma _{B}$ and $\gamma _{M}$ .

\section{Conclusion}

We do not consider here the experimental situation, which is now unclear an
controversial even for the $SF$ hybrid structures with thick ferromagnetic
layers. Let us only note, that spin-orbit scattering is relevant for
ferromagnetic conductors containing large $Z$ number. Magnetic inhomogeneity
of the materials, such as the $F$ layer multidomain structure, domain walls,
inhomogeneous ''cryptoferromagnetic'' state imposed by superconductor also
give nonzero probability amplitude for the spin-flip scattering. For
nanoscale hybrid structures strong electric field arising near metal-metal
boundaries is also an important source of the spin-flip processes. In the
absence of any precise information about the magnetic structure of the
samples used in experiments on $SF$ sandwiches, we restrict ourselves to
making only qualitative calculation.

In conclusion, we investigate the spin-orbit scattering effect on critical
current of $SFIFS$ tunnel junction for the case of thin $F$ layers, when the
superconducting order parameter oscillation is absent. Instead, the
parameter phase jumps at the $SF$ interfaces. The analysis is based on a
microscopic theory for proximity coupled $S$ and $F$ layers. The main result
of our calculations is that the spin-flip processes can sizably modify the
behavior of the $dc$ Josephson current versus the exchange field for $SFIFS$%
\ tunnel junction. We find that the reduction of the exchange field effects
is nonlinear in character getting its maximum in the ferromagnetic field's
region where the critical current enhancement or the transition to the $\pi $%
-state occur. Hence, for understanding the various experimental results on
tunnel structures with thin $F$ layers, the coupled effects of the exchange
interaction and spin-orbit scattering must be considered.

The authors are grateful to Valery Ryazanov and Elena Koshina for useful
discussions.

\begin{center}
Figure Captures
\end{center}

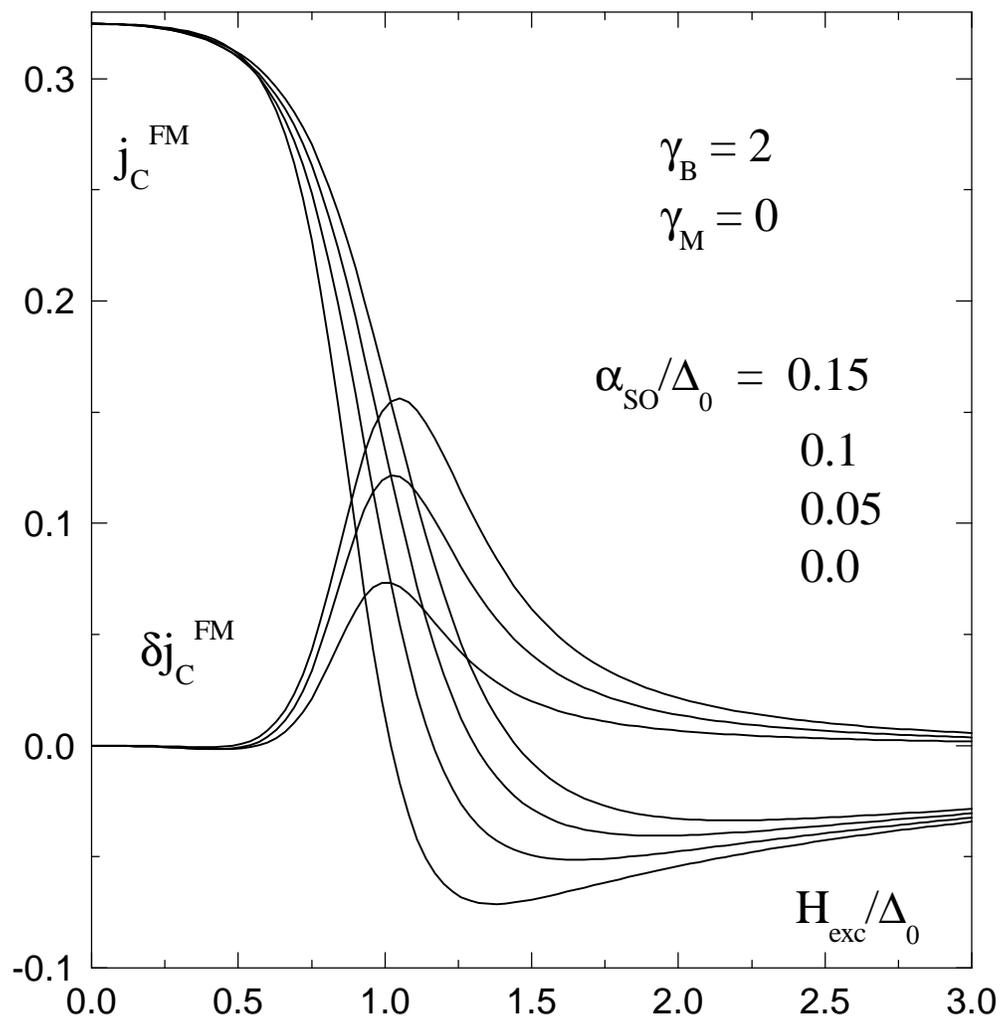
Fig. 1. \ Critical current of $SFIFS$ tunnel junction with parallel
orientation of the $F$ layers magnetization $j_{C}^{FM}$ versus exchange
energy for various $\alpha _{SO}/\Delta _{0}$ = 0, 0.05 , 0.1 and 0.15\ ;\
the $SF$ interface transparency is low $\gamma _{B}=2$ and proximity effect
is weak $\gamma _{M}$ = 0. The additive part of critical current due to
spin-orbit scattering, $\delta j_{C}^{FM}$ , is also shown. $T=0.1T_{C}$ .

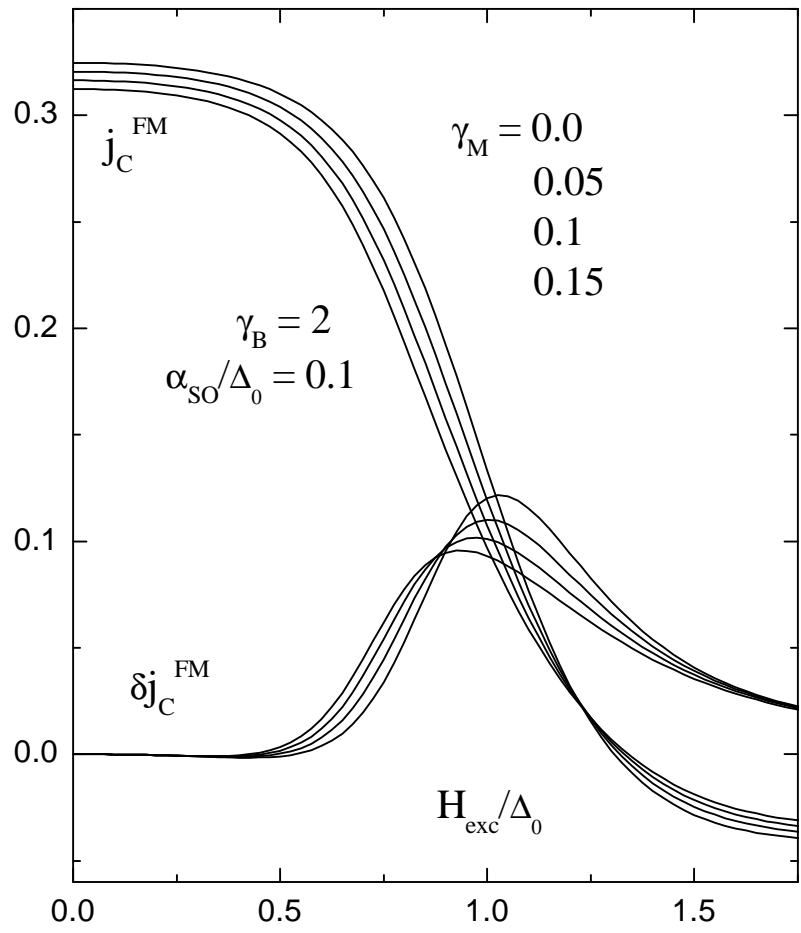
Fig. 2. \ Critical current of $SFIFS$ tunnel junction with parallel
orientation of the $F$ layers magnetization $j_{C}^{FM}$ versus exchange
energy for various $\gamma _{M}$ = 0, 0.05, 0.1, 0.15; $\gamma _{B}=2$ and $%
\alpha _{SO}/\Delta _{0}$ = 0.1. The additive part of critical current due
to spin-orbit scattering, $\delta j_{C}^{FM}$ , is also shown. $T=0.1T_{C}$ .

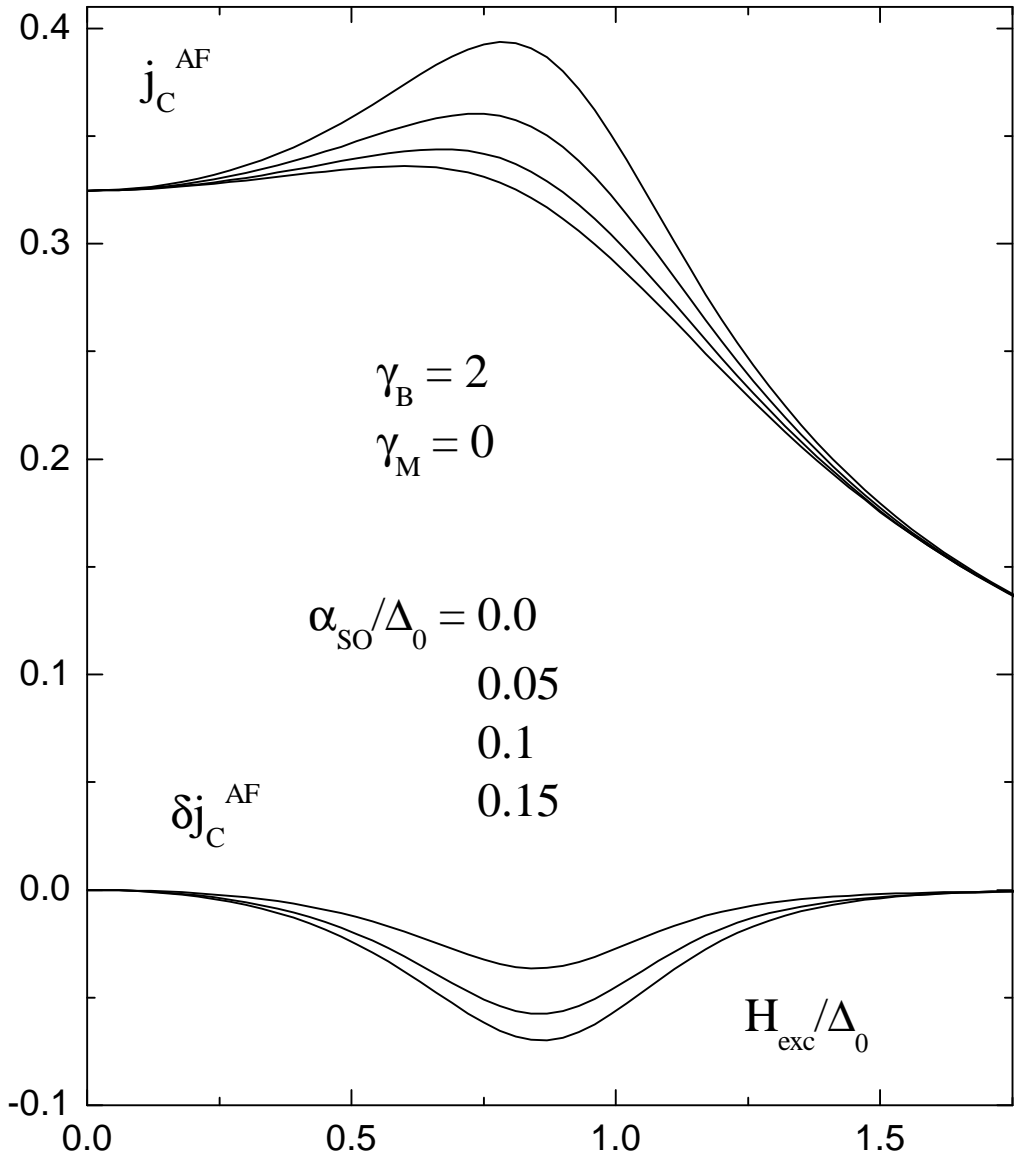
Fig. 3. \ Same as on Fig.1 but for antiparallel orientation of the $F$
layers magnetization. The additive part of critical current due to
spin-orbit scattering, $\delta j_{C}^{AF}$ , is also shown.

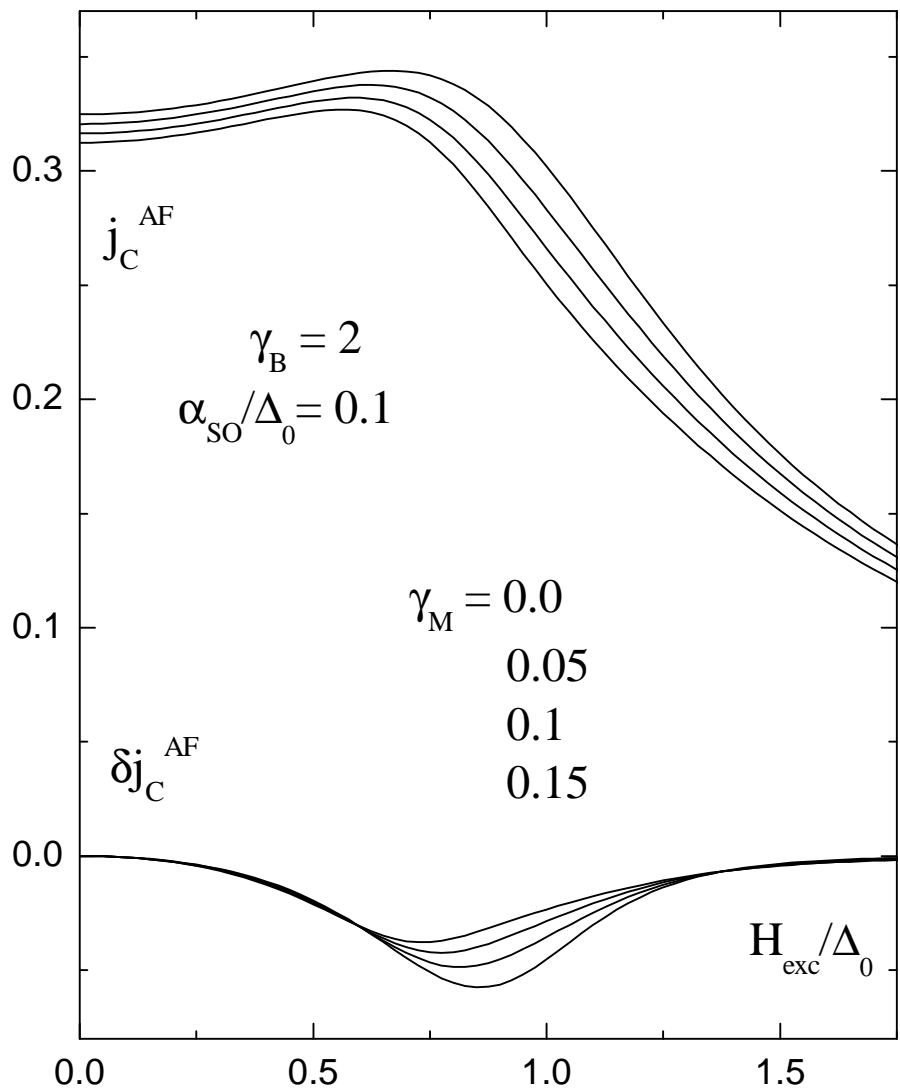
Fig. 4. \ Same as on Fig.2 but for parallel orientation of the $F$ layers
magnetization. The additive part of critical current due to spin-orbit
scattering, $\delta j_{C}^{FM}$ , is also shown.

\end{document}